\documentclass[aps,prl,twocolumn,superscriptaddress,amsmath,amssymb]{revtex4}
\usepackage{float}
\usepackage{makeidx}
\usepackage{graphicx,amsmath}
\usepackage{CJK}
\usepackage{colordvi}
\usepackage{color}

\begin{document}
\begin{CJK*}{GB}{gbsn}

\title{Photoassociation spectroscopy of weakly bound $^{87}Rb_{2}$ molecules near $5P_{1/2} +5S_{1/2}$ threshold by optical Bragg scattering in Bose-Einstein condensates}

\author{Khan Sadiq Nawaz}
\affiliation{State Key Laboratory of Quantum Optics and Quantum
Optics Devices, Institute of Opto-Electronics, Collaborative Innovation Center of Extreme Optics, Shanxi University,
Taiyuan 030006, P.R.China }

\author{Liangchao Chen$^*$}

\affiliation{State Key Laboratory of Quantum Optics and Quantum
Optics Devices, Institute of Opto-Electronics, Collaborative Innovation Center of Extreme Optics, Shanxi University,
Taiyuan 030006, P.R.China }

\author{Chengdong Mi}

\affiliation{State Key Laboratory of Quantum Optics and Quantum
Optics Devices, Institute of Opto-Electronics, Collaborative Innovation Center of Extreme Optics, Shanxi University,
Taiyuan 030006, P.R.China }

\author{Zengming Meng}

\affiliation{State Key Laboratory of Quantum Optics and Quantum
Optics Devices, Institute of Opto-Electronics, Collaborative Innovation Center of Extreme Optics, Shanxi University,
Taiyuan 030006, P.R.China }

\author{Lianghui Huang}

\affiliation{State Key Laboratory of Quantum Optics and Quantum
Optics Devices, Institute of Opto-Electronics, Collaborative Innovation Center of Extreme Optics, Shanxi University,
Taiyuan 030006, P.R.China }

\author{Pengjun Wang}

\affiliation{State Key Laboratory of Quantum Optics and Quantum
Optics Devices, Institute of Opto-Electronics, Collaborative Innovation Center of Extreme Optics, Shanxi University,
Taiyuan 030006, P.R.China }

\author{Jing Zhang$^{\dagger}$}

\affiliation{State Key Laboratory of Quantum Optics and Quantum
Optics Devices, Institute of Opto-Electronics, Collaborative Innovation Center of Extreme Optics, Shanxi University,
Taiyuan 030006, P.R.China }

\begin{abstract}
We report the high resolution photoassociation (PA) spectroscopy
of a $^{87}Rb$ Bose-Einstein condensate (BEC) to excited molecular
states near the dissociation limit of $5P_{1/2} +5S_{1/2}$ by
optical Bragg scattering. Since the detection of optical Bragg
scattering in BEC has a high signal-noise ratio, we obtain the
high resolution PA spectrum of excited molecular states in the
range of $\pm1$ GHz near the dissociation limit of $5P_{1/2}
+5S_{1/2}$. We compare the results with the conventional method of
trap loss and show that the results agree each other very well.
Many interesting phenomena of excited molecular states are
observed, such as light-induced frequency shift and the anomalous
strong bound molecular lines at the atomic transition from
$|F=1\rangle$ to $|F^{\prime}=2\rangle$. The observed excited
molecular states in the range of $\pm1$ GHz near the dissociation
limit of $5P_{1/2} +5S_{1/2}$ are never reported before, which
will help to further improve the long range bound state models
near the dissociation limit.

\end{abstract}
\maketitle
\end{CJK*}

The photoassociation (PA) process of cold
atoms~\cite{Stallway1999,Jones2006} has become one of the most
precise techniques for spectroscopy of long range molecular
states, which is very important for studies of cold molecules,
atom-atom collisions, atom-molecule collisions, etc. The desire to
achieve quantum degeneracy in molecules was a natural consequence
after the first observation of the BEC in alkali
atoms~\cite{Anderson1995,Davis1995,Bradley1995}. One of the
approaches is the indirect production of ultracold molecules by
associating atoms to form molecules from cooled
atoms~\cite{Ni2008,Danzl2008,Molony2014,Park2015,Guo2016,elberg2018}.
The PA process is the first step of this excitation-deexcitation
scheme~\cite{Ni2008,Danzl2008,Molony2014,Park2015,Guo2016,elberg2018}.
Due to this reason, PA spectroscopy in ultracold atoms must be
studied first, which provides accurate and useful data that could
determine or predict new routes in cold-molecule formation.
Moreover, the PA process has become an important tool for tuning
interatomic
interactions~\cite{Fedichev1996,Bohn1999,Enomoto2008,Yamazaki2010,Blatt2011,Yan2013,Junker2008,Bauer2009,Fu2013,Peng2018},
called optical Feshbach resonance, in which free atom pairs are
coupled to an excited molecular state by laser field near a PA
resonance.

The PA spectroscopy usually employs the trap loss detection
technique~\cite{Wang1996,Gerton2001,McKenzie2002} or two-photon
ionization~\cite{Gabbanini2000}. The trap loss detection consists
of recording the trap fluorescence (proportional to the atom
number) for cold atoms in magneto-optical trap
(MOT)~\cite{Wang1996} or measuring the remaining atoms for quantum
degenerate atoms in the trap~\cite{Gerton2001,McKenzie2002} while
the PA laser is frequency scanned. If the laser wavelength is
resonant with a molecular state, the PA will create long-range
excited molecules. These excited molecules, having a very short
lifetime will rapidly decay, inducing atom loss from the MOT or
trap. In this paper, we develop a method of optical Bragg
scattering in $^{87}Rb$ BEC to measure PA spectrum, which is
especially suitable for near the dissociation limit (near the
atomic resonant transition). Optical Bragg scattering from optical
lattice is a widely used method for observing and analyzing
periodic structures, in which light is incident on the atomic
layers confined by optical standing wave at well defined angle and
is Bragg reflected (scattered). It has been studied experimentally
in cold atomic
gases~\cite{Birkl1995,Weidemuller1996,Slama2005,Hart2015,Chen2018}.
The signal of optical Bragg scattering in BEC is strong when the
probe light is near the atomic resonant transition. Therefore, we
obtain the high resolution photoassociation spectrum of excited
molecular states in the range of $\pm1$ GHz near the dissociation
limit of $5P_{1/2} +5S_{1/2}$ of $^{87}Rb$ atoms.

The PA spectra of $^{87}$Rb$_{2}$ $5P+5S$ excited molecular bound
states have been studied and analysed in great detail in
\cite{Hoffmann1992, Hoffinann1994, Peters1994, Cline1994,
Fioretti2001, Gutterres2002, Kemmann2004, Jelassi2006,
Jelassi2006-1,Hamley2009,Tsai2013,Jelassi2014}. In this paper we
study the $^{87}$Rb$_{2}$ weakly bound excited dimers formation in
an $^{87}$Rb BEC using PA within $\pm1$ GHz of the D1 hyperfine
asymptotes, which is never reported before. Previous studies in
this region could not detect the quantized losses of PA for atoms
in MOT and only a
continuous loss was observed \cite{Peters1994}. 
The quantum degenerate gases bring several advantages for PA
spectroscopy, such as the PA rate is enhanced since it increases
proportionally with phase space density and the spectroscopic
precision is increased since the energy spread of atoms at low
temperature is low. We also compare the PA spectra with the
conventional method of trap atom loss and show the mutual
agreement. Many interesting phenomena of excited molecular states
are observed, such as light-induced frequency red shift, the
simple ordered molecular lines, and the anomalously strong bound
molecular lines at the atomic transition from $|F=1\rangle$ to
$|F^{\prime}=2\rangle$.

\begin{figure}
\centerline{\includegraphics[width=8cm]{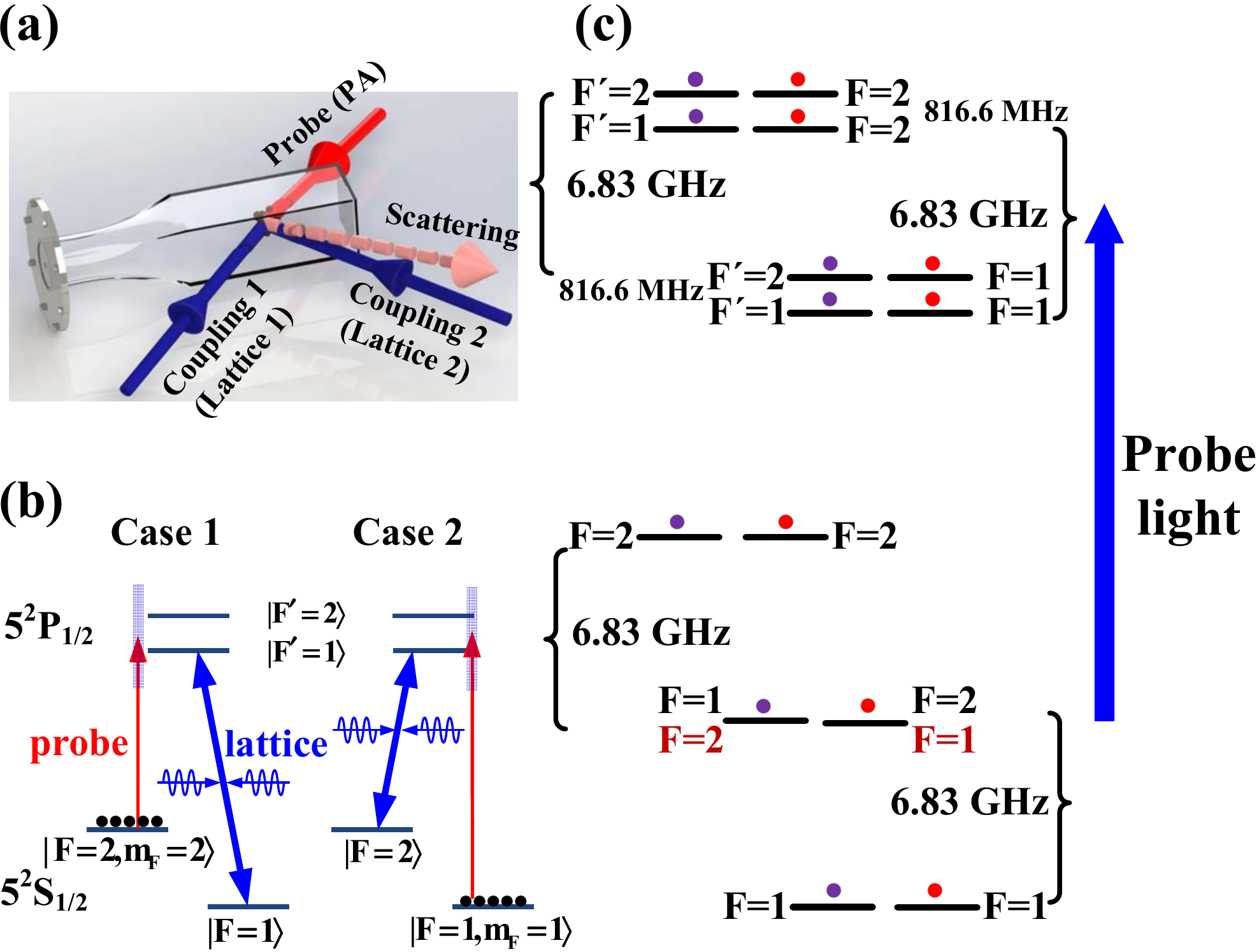}} \vspace{0.06in}
\caption{\textbf{Schematics of laser configuration and the energy levels.} (a) The laser configuration for the experiment. A pair of strong coupling laser beams forming (around 795 nm) one-dimensional optical lattice. The probe (PA) laser illuminate BEC to generate Bragg emission. (b) Energy diagram of the $5^{2}S_{1/2}-5^{2}P_{1/2}$ transitions of $^{87}$Rb. Case 1: A pair of strong coupling laser drive the transition between
$|F=1\rangle$ and $|F'=1\rangle$. Atoms are prepared in the
$|F=2,m_{F}=2\rangle$ state and the frequencies of the probe (PA) laser are scanned below or above the lower $|F^{\prime}=1\rangle$ and upper excited hyperfine level $|F^{\prime}=2\rangle$ of the D1 line. Case 2: The coupling lasers drive the transition between
$|F=2\rangle$ and $|F'=1\rangle$ and atoms are prepared in the
$|F=1,m_{F}=1\rangle$ state. (c) Energy diagram of the dissociation limit of the ground and excited bound molecular state.  \label{Fig1} }
\end{figure}

The experimental setup and energy levels are shown in Fig. 1. The starting point for our experiments is an essentially pure BEC
with typically $5\times10^5$ $^{87}Rb$ atoms in the
$|F=2,m_{F}=2\rangle$ hyperfine ground state sublevel confined in a
cross-beam dipole trap at a wavelength near 1064 nm \cite{Chen2018}. The geometric
mean of trapping frequencies $\overline{\omega}\simeq2\pi\times80$
Hz in our system. The atoms can be transferred to
the $|F=1,m_{F}=1\rangle$ state via a rapid adiabatic
passage induced by a microwave frequency field at 2 G of the bias magnetic field.

The optical Bragg scattering technique uses a weak probe (PA)
laser and a pair of strong coupling laser beams forming
one-dimensional optical lattice as shown in Fig. 1(a). Both of the
coupling beams are derived from the same laser and they intersect
at an angle of $48^{0}$. The coupling laser beams have a waist
($1/e^{2}$ radius) of about 280 $\mu m$ at the position of the
BEC. The weak probe laser acting as PA light has a waist of about
600 $\mu m$. The relative frequencies of the lattice and probe
laser are locked by the optical phase locked loop (OPLL). The
probe (PA) laser frequency is changed by setting the frequency of
local oscillator of the OPLL. The intersecting angle between the
probe light and the optical Bragg emission is about $132^{0}$,
which satisfy the phase matching condition as discussed in our
previous work \cite{Chen2018, wang2020}. In order to obtain the
dark background and high signal-noise ratio for detecting the
Bragg emission, the intersecting angle between the plane of the
two coupling beams and the plane of the probe-Bragg beams is kept
at $11^{0}$.

When atoms are prepared in the $|F=2,m_{F}=2\rangle$ state, the
frequency of the coupling laser beams is locked at transition from
$|F=1\rangle$ to $|F^{\prime}=1\rangle$ as shown in Fig. 1(b). At
the same time, the frequency of the probe (PA) laser is scanned to
near the transition between $|F=2,m_{F}=2\rangle$ and the excited
state $|F^{\prime}=2\rangle$. In this configuration, the coupling
beams form a far detuning optical lattice that mainly modulate the
initial state $|F=2,m_{F}=2\rangle$, such that the refractive
index of the probe beam is periodically modulated and a structure
similar to a photonic crystal is formed. Therefore, the
directional emission generated in this configuration corresponds
to the ordinary optical Bragg scattering. The spectrum of optical
Bragg scattering can be obtained by scanning the frequency of the
probe (PA) laser and fixing the frequency of the coupling beams.
However, when the probe (PA) laser drive atoms from the transition
from $|F=2,m_{F}=2\rangle$ and the excited state
$|F^{\prime}=1\rangle$, a pair of the coupling laser beams
together with the probe laser form standing wave-coupled
electromagnetically induced transparency (EIT) configuration. This
case becomes the special form of optical Bragg scattering, called
superradiance scattering \cite{Wang2015}. The lineshape of the
ordinary optical Bragg scattering spectra is different from that
of superradiance scattering \cite{Chen2018}. Here we study two
cases as shown in Fig. 1(b). \textbf{Case 1:} Atoms are prepared
in the $|F=2,m_{F}=2\rangle$ state, the frequency of the coupling
laser beams is locked at transition from $|F=1\rangle$ to
$|F^{\prime}=1\rangle$ (or $|F^{\prime}=2\rangle$); \textbf{Case
2:} Atoms are prepared in the $|F=1,m_{F}=1\rangle$ state, the
frequency of the coupling laser beams is locked at transition from
$|F=2\rangle$ to $|F^{\prime}=1\rangle$ (or
$|F^{\prime}=2\rangle$). In the optical Bragg scattering spectra,
the strength of Bragg scattering will be reduced when the probe
laser is resonant with an excited molecular state due to atom
losses of BEC. By this process, we obtain the spectrum of excited
molecular states.

\begin{figure}
\centerline{\includegraphics[width=8cm]{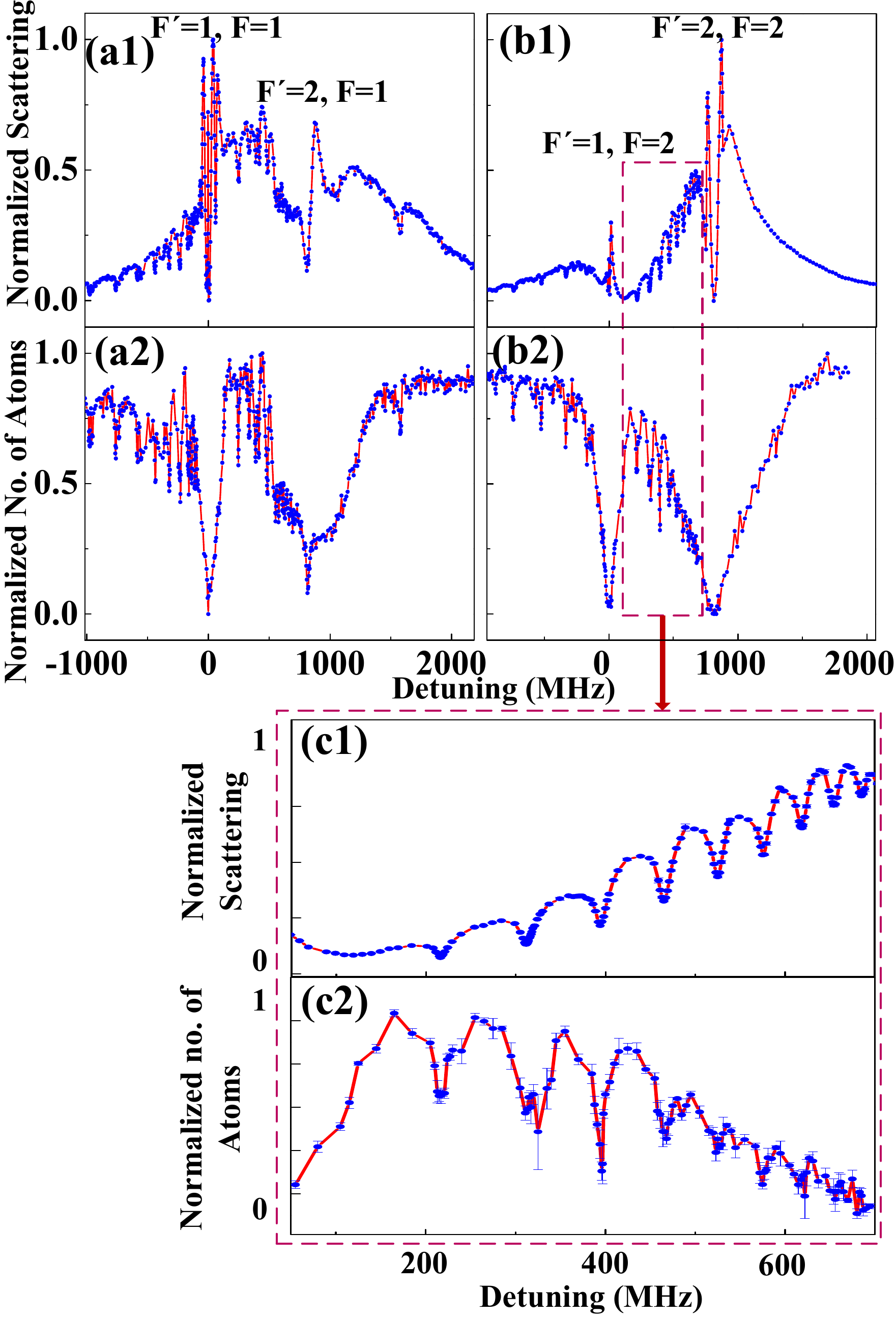}} \vspace{0.06in}
\caption{\textbf{PA spectra obtained by optical Bragg scattering
and trap loss respectively.} (a1) and (b1) PA spectra are measured
by optical Bragg scattering when BEC is prepared in the $|F=1,
m_{F}=1\rangle$ and $|F=2, m_{F}=2\rangle$ state respectively. The
frequency of the coupling laser is locked at transition from
$|F=2\rangle$ to $|F^{\prime}=1\rangle$ and from $|F=1\rangle$ to
$|F^{\prime}=2\rangle$ respectively. The power of each coupling
laser beam is 200 $\mu$W for(a1) and 400 $\mu$W for(b1) while the
power of the probe laser is 25 $\mu$W. (a2) and (b2) PA spectra
are measured by trap loss without the coupling laser beams, when
BEC is prepared in the $|F=1, m_{F}=1\rangle$ and $|F=2,
m_{F}=2\rangle$ state respectively. The power of the probe laser
is 25 $\mu$W. (c1) and (c2) are enlarged plots of (b1) and (b2),
in which error bars are given. Every point is recorded three times
and the average points are connected by a line. \label{Fig2} }
\end{figure}

To get the Bragg scattering spectra, all the optical fields,
including the coupling and probe (PA) lasers, illuminate BEC in
the optical dipole trap simultaneously for 20 $\mu$s, and
simultaneously the resulting Bragg emission is detected by EMCCD.
We also obtain PA spectra by trap loss, in which the remaining
atoms are measured by the time of flight absorption imaging after
exposing the BEC to this single probe (PA) laser for 20 $\mu$s.
The difference between the two techniques is that in the Bragg
scattering two coupling laser beams makes a density modulated
lattice in the BEC and the probe (PA) laser scattered from this
lattice, is recorded live by the EMCCD while in the trap loss
method, only the probe (PA) laser interacts with the BEC and then
the BEC is left to expand for 30 ms after which an image is taken
by imaging laser to count the number of remaining atoms. Fig. 2
shows the optical Bragg scattering and trap loss spectra of the
$^{87}$Rb atoms prepared in the $|F=1, m_{F}=1\rangle$ and $|F=2,
m_{F}=2\rangle$ states respectively, when the frequencies of the
probe (PA) laser are scanned below or above the lower
$|F^{\prime}=1\rangle$ and upper excited hyperfine level
$|F^{\prime}=2\rangle$ of the D1 line. Here, the frequency of the
coupling laser beams is locked at transition from $|F=2\rangle$ to
$|F^{\prime}=1\rangle$ (Fig. 2(a1)) when atoms are initially
prepared in $|F=1, m_{F}=1\rangle$ and from $|F=1\rangle$ to
$|F^{\prime}=2\rangle$ (Fig. 2(b1)) when atoms are prepared in
$|F=2, m_{F}=2\rangle$. The spectrum of optical Bragg scattering
presents the broader peaks on both sides of the atomic resonance
(marked by the respective atomic hyperfine quantum numbers) and
several narrower and shallower dips with the reduced scattering
corresponding to the weak bound molecular states due to atomic
loss by PA. As a comparison, Fig. 2(a2) and (b2) show PA spectra
measured by trap loss without the coupling laser beams (no
lattice, only BEC+probe (PA) laser), when BEC is prepared in the
$|F=1, m_{F}=1\rangle$ and $|F=2, m_{F}=2\rangle$ state
respectively. The lineshape of trap loss spectrum shows broader
Lorentzian dip near the atomic resonance and several narrower and
shallower dips corresponding to the weakly bound molecular states.
Fig. 2(c1) and (c2) show the enlarged plots of Fig. 2(b1) and
(b2). From the error bars, PA spectrum by optical Bragg scattering
detection present the high signal-noise ratio compared with that
of trap loss especially in the region of near dissociation
threshold. One reason is that the spectrum of optical Bragg
scattering presents the broad M lineshape (dip at the resonance)
\cite{Schilke2015}, in contrast, the lineshape of trap loss
presents broad Lorentzian dip near the atomic resonance. The other
is that the spectrum for trap loss is prone to more noise compared
to the spectrum for optical Bragg scattering. The trap loss
technique employs the absorption images to count the number of
remaining atoms. The absorption images have several technical
noise sources, such as the interference fringes and the change in
imaging laser power between recording of the absorption and the
reference image \cite{Kristensen2019}, while the Bragg sattering
technique measures the scattered light live from the dark
background during the scattering process and avoid these extra
steps. Our results demonstrate that the two methods compliment
each other, however, the Bragg scattering technique has a high
signal to noise ratio near atomic resonance.

\begin{figure}
\centerline{\includegraphics[width=8cm]{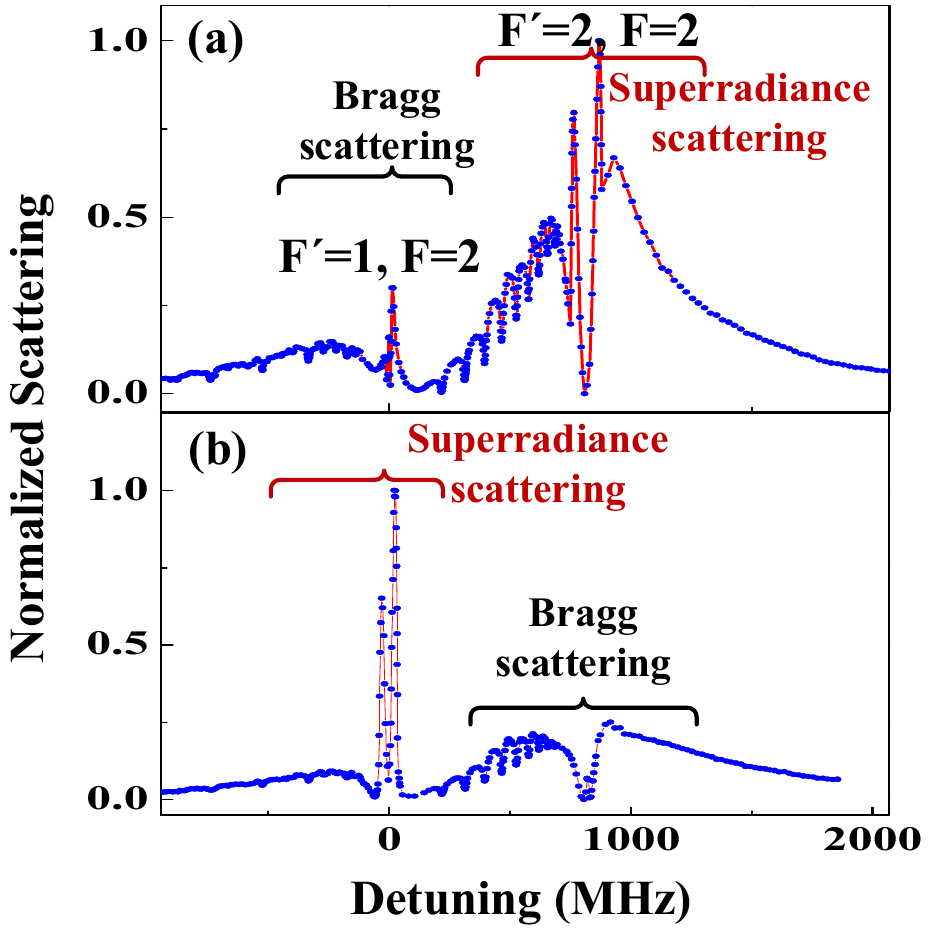}} \vspace{0.06in}
\caption{\textbf{Comparison of spectra for optical Bragg
scattering and superradiance scattering.} The frequency of the
coupling laser beams is locked at transition from $|F=1\rangle$ to
$|F^{\prime}=2\rangle$ (a) and from $|F=1\rangle$ to
$|F^{\prime}=1\rangle$ (b), respectively. BEC is prepared in the
$|F=2, m_{F}=2\rangle$ state.  \label{Fig4} }
\end{figure}

The spectral lineshapes of optical Bragg scattering and
superradiance scattering have been studied in our previous work
\cite{Chen2018}. In the case of Bragg scattering, the coupling
beams forms a far detuned lattice which modulate the ground state
BEC to generate a periodic structure in the ground state atoms,
which induce a directional optical scattering. The spectrum of
optical Bragg scattering presents a broader M lineshape, which is
clearly shown in the Bragg scattering part of Fig. 3(b) when the
frequencies of the probe laser are scanned cross an atomic
transition. For the superradiance scattering, the coupling laser
and the probe (PA) laser couple the same excited state to
different ground states to induce alternate periodic structures of
atoms in the excited state and ground state \cite{Chen2018,
wang2020}. The superradiance scattering spectrum present two
narrow and intense peaks at the two side of resonance (the
superradiance scattering part of Fig. 3(a)) due to the density of
state of the periodic structure in the excited state. We can see
that the spectral lineshapes in Fig. 3 are different for the probe
transitions near the $|F\rangle\rightarrow|F^{\prime}=F-1\rangle$,
$|F\rangle\rightarrow|F^{\prime}=F\rangle$, and
$|F\rangle\rightarrow|F^{\prime}=F+1\rangle$. Here, both cases
have the same broad wings. Therefore, the lineshape of
superradiance and Bragg scattering are quite different near the
atomic resonance,  however, this lineshape does not influence the
weak bound molecular states since the PA loss is measured using
these broad wings.

\begin{figure}
\centerline{\includegraphics[width=8cm]{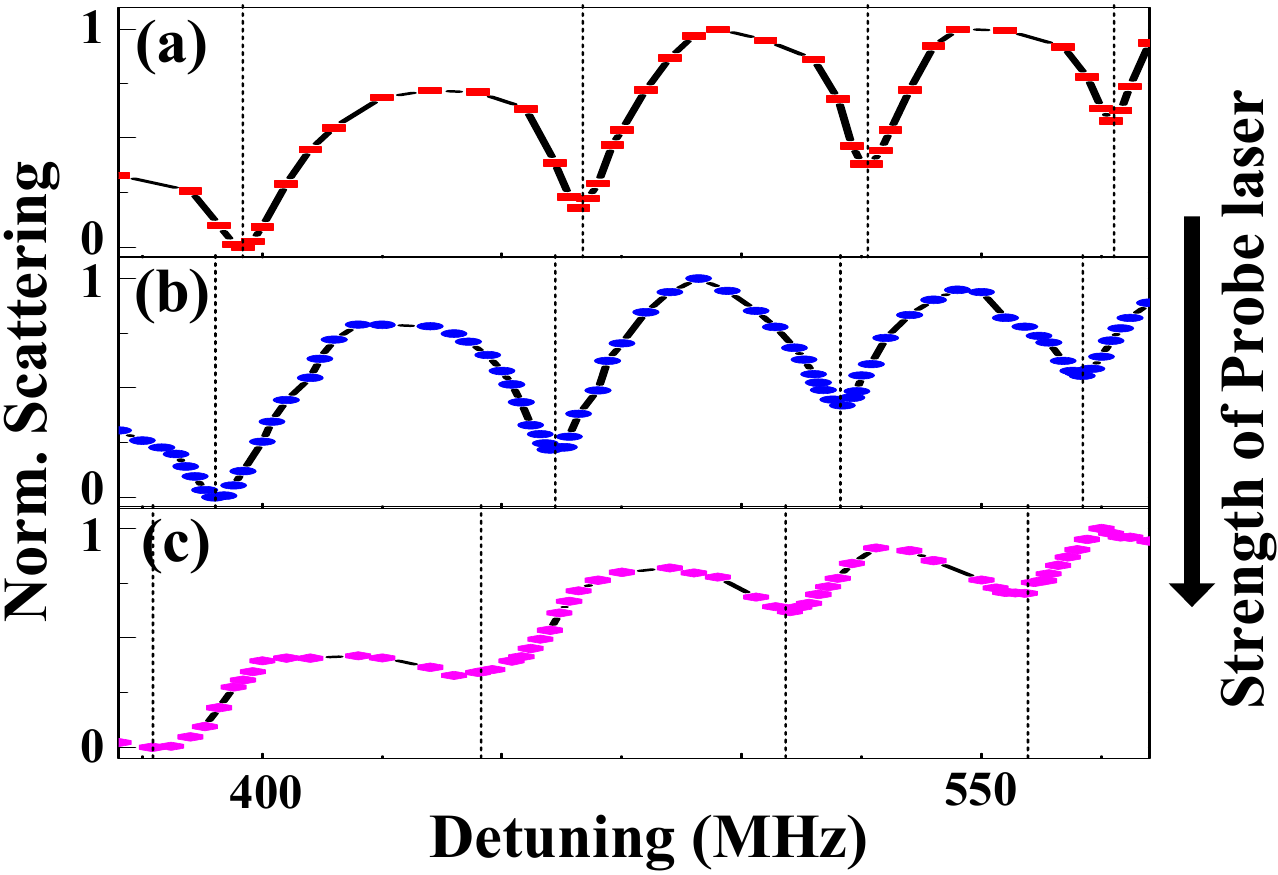}} \vspace{0.06in}
\caption{\textbf{PA frequency red shift.} The power of PA laser is 25 $\mu$W (a), 100 $\mu$W (b), 300 $\mu$W (c), respectively. \label{Fig6} }
\end{figure}

PA light-induced frequency red shift is a typical characteristic
of PA spectra \cite{Gerton2001,McKenzie2002}. We also observe this
phenomenon in the PA spectra where the lines shift to the red with
increasing PA laser power as shown in Fig. 4. This phenomenon
arises from coupling of the various ground-state threshold
scattering states to the excited-state bound molecular levels by
PA light. The density of ground continuum states always increases
versus energy and therefore, the bound states shift in frequency
with the addition of new ground states from the continuum.

We present the observed PA lines in tabular form as shown in Table
I. Those PA lines are presented which are observable in both of
the techniques. The lines are recorded for a probe (PA) laser
power of 25 $\mu$W. We use the same probe power for the two
techniques in order to avoid PA light-induced frequency red shift.
Thus, we obtain mutually consistent molecular dips using the two
techniques for a large number of peaks with an uncertainty of
$\pm$0.2 MHz.

Now we analyze the weak bound molecular states near the
dissociation limit, which we have divided into several regions in
Table I, as follows. \textbf{Region A}: There are no weakly bound
molecular states due to being higher than the last dissociation
limit ($F^{\prime}=2,F=2$). Here, the $5P_{3/2} +5S_{1/2}$
molecular threshold is not considered due to far blue detuning of
15 nm. \textbf{Region B}: The spectrum of the weakly bound
molecular states represent the simple ordered lines of the
vibrational levels of single molecular potential curve. Here, the
rotational energy is neglected due to the huge internuclear
separation (the rotational constant has a magnitude on the order
of only several MHz) \cite{Kemmann2004}. So the rotational
structure is unresolved in our work. \textbf{Regime C}: The bound
molecular lines are not very regular, since there are two
molecular potential energy curves (one terminating at
$F^{\prime}=1$ and the other at $F^{\prime}=2$) and thus two
vibrational series are overlapped together. \textbf{Region D}:
There should be no the bound molecular states in this regime due
to blue detuning of the dissociation limit ($F^{\prime}=2,F=1$).
However, one bound molecular line is observed, which originate
from the molecular potentials of the higher dissociation limit
($F^{\prime}=1,F=2$) and ($F^{\prime}=2,F=2$). The reason is
explained as follow in Region E. \textbf{Region E}: The coupling
strength (Franck-Condon factor) between the ground continuum state
and the excited-state bound molecular state in this region are
anomalously stronger than that of other regions at the same power
of PA laser. The width of each bound molecular line is determined
by the coupling strength between the ground continuum state and to
the excited-state bound molecular state. Therefore, the power
broadening of the bound molecular lines in this region causes the
lineshape of the atomic transition from $|F=1\rangle$ to
$|F^{\prime}=2\rangle$ to deviate from a standard Lorentzian
shape, as shown in Fig. 5. Upon reducing the power of PA laser, we
can see that many narrow bound molecular lines appear in red
detuning side at the atomic resonant transition (from
$|F=1\rangle$ to $|F^{\prime}=2\rangle$) and the whole lineshape
of the atomic transition recovers to the broad Lorentzian shape.
This anomalous feature may be explained in terms of an additional
broadening of the resonance due to the strong coupling with the
deep bound molecular channels of ($F^{\prime}=1,F=2$) and
($F^{\prime}=2,F=2$). \textbf{Region F}: The bound molecular lines
are not very regular, which is similar to Regime III.

\begin{figure}
\centerline{\includegraphics[width=8cm]{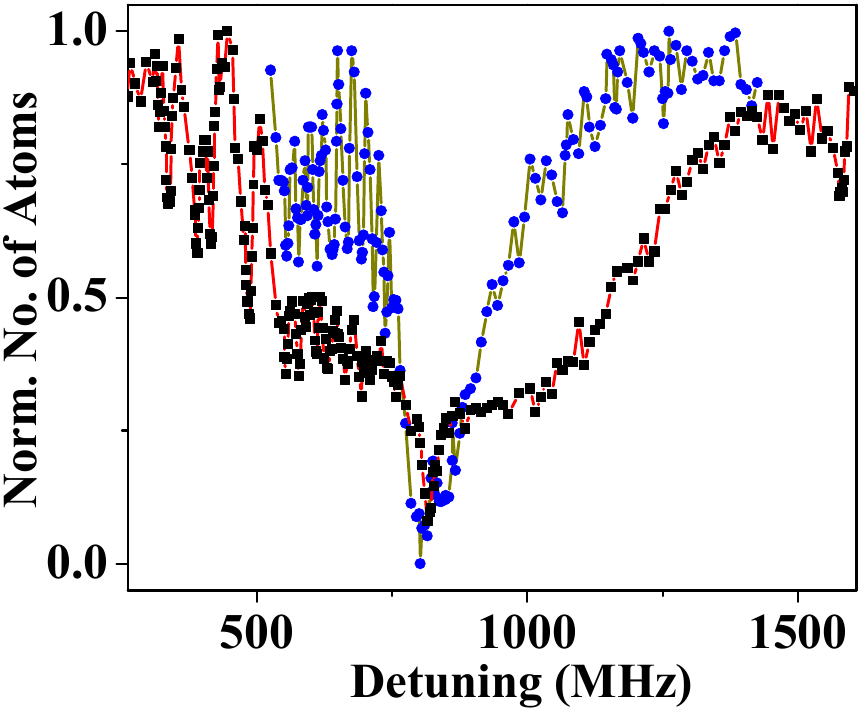}} \vspace{0.06in}
\caption{\textbf{Dependence of the lineshape on the probe power
using the trap loss technique at the probe frequency resonant to
the atomic transition from $|F=1\rangle$ to
$|F^{\prime}=2\rangle$.} The PA laser power is 25~$\mu$W (squares
with red line) and 3~$\mu$W (circular with gray line).
\label{Fig8} }
\end{figure}

\begin{table}
\centerline{\includegraphics[width=8cm]{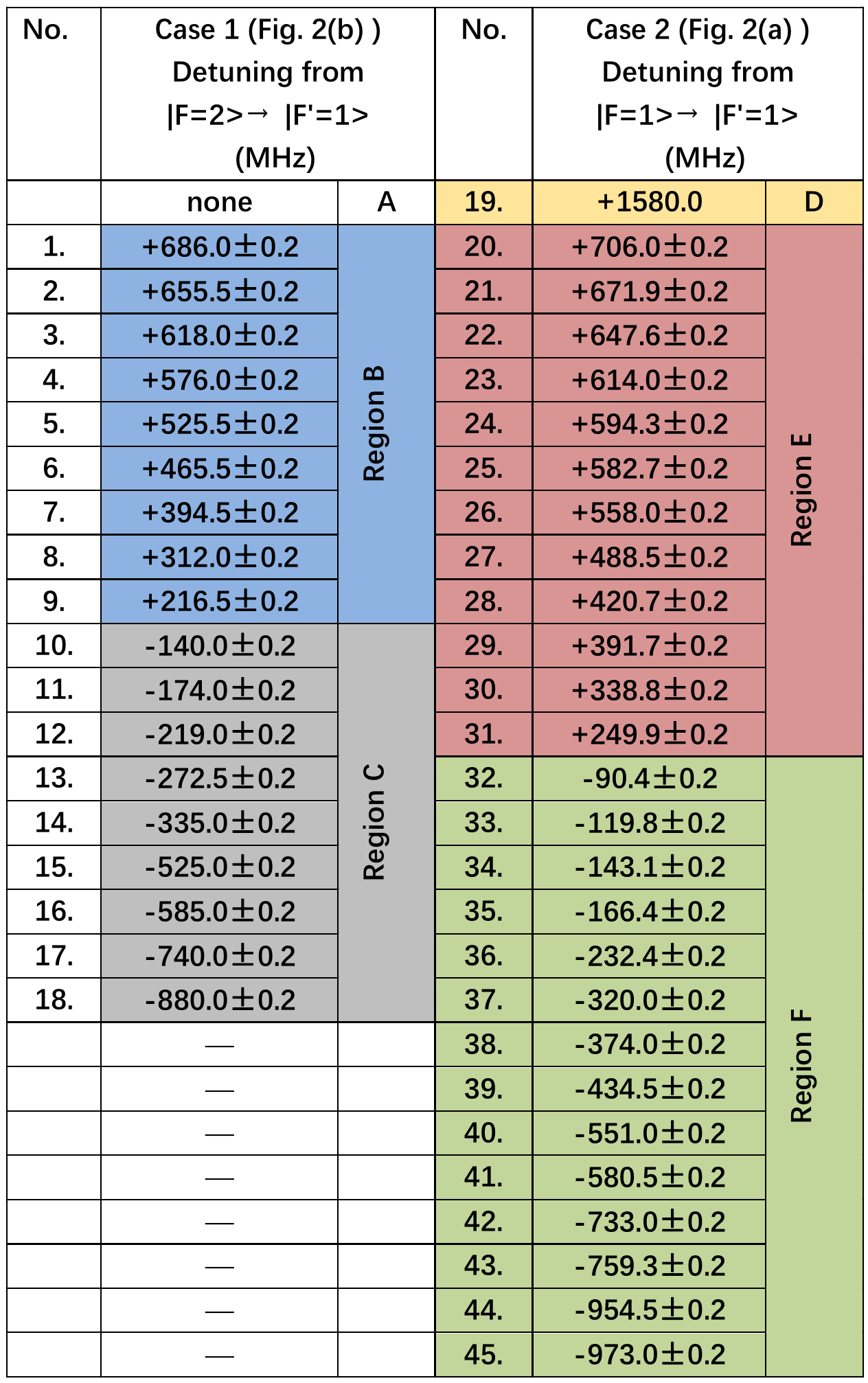}} \vspace{0.06in}
\caption{The observed positions of the PA resonances measured by
their detuning from $|F=2\rangle\rightarrow|F^{\prime}=1\rangle$
transition (Case 1) and
$|F=1\rangle\rightarrow|F^{\prime}=1\rangle$ transition (Case 2)
of the D1 line of Rb. The colors label the different frequency
regions, where region A is the frequency range of blue detuning
$|F=2\rangle\rightarrow|F^{\prime}=2\rangle$ transition in Case 1,
region B is between $|F=2\rangle\rightarrow|F^{\prime}=1\rangle$
and $|F=2\rangle\rightarrow|F^{\prime}=2\rangle$ in Case 1, region
C is red detuning $|F=2\rangle\rightarrow|F^{\prime}=1\rangle$ in
Case 1, region D is blue detuning
$|F=1\rangle\rightarrow|F^{\prime}=2\rangle$ transition in Case 2,
region E is between $|F=1\rangle\rightarrow|F^{\prime}=1\rangle$
and $|F=1\rangle\rightarrow|F^{\prime}=2\rangle$ in Case 2, region
F is red detuning $|F=1\rangle\rightarrow|F^{\prime}=1\rangle$ in
Case 2. \label{TabI} }
\end{table}

In conclusion, we demonstrate high resolution PA spectroscopy of
weakly bound $^{87}Rb_{2}$ molecular near $5P_{1/2} +5S_{1/2}$
threshold by optical Bragg scattering in Bose-Einstein
condensates. Optical Bragg scattering provides a useful tool to
measure the PA spectra near the atomic resonant transition. We
present the high resolution PA spectrum of excited molecular
states in the range of $\pm1$ GHz near the dissociation limit of
$5P_{1/2} +5S_{1/2}$ of $^{87}Rb$ atoms, which is never studied in
this region before. For comparison, we also perform the
measurement with the traditional method of trap atom loss and the
strong consistency between the two methods provides confirmation
for its validity. The last few bound molecular states from the
dissociation limit are of great interest as this can improve the
long range molecular potential curves. In the future, combined
with the theoretical calculations about long range molecular
potential curves near $5P_{1/2} +5S_{1/2}$ threshold, we can
further supplement and improve the data of PA spectra including
the finer and smaller molecular transition lines. Moreover due to
anomalous strong bound molecular lines at the atomic transition
from $|F=1\rangle$ to $|F^{\prime}=2\rangle$, the contribution of
bound excited molecular states to the lineshape of the atomic
resonant transition will play some role, when using this atomic
resonant transition spectrum to study precise spectroscopic
phenomena, such as EIT phenomena.

$^*$ chenlchao@gmail.com.
$^{\dagger}$ jzhang74@sxu.edu.cn,
jzhang74@yahoo.com.

\begin{acknowledgments}
This research is supported by the MOST (Grant No. 2016YFA0301602),
NSFC (Grant No. 11234008, 11474188, 11704234) and the Fund for
Shanxi "1331 Project" Key Subjects Construction.

\end{acknowledgments}

\end{document}